\documentclass[useAMS,usenatbib,usegraphicx]{mn2e}

\title[H~I opacity of the IGM]{The H~I opacity of the intergalactic medium at redshifts $1.6 < z < 3.2$}
\author[D. Kirkman \etal]{David Kirkman\thanks{E-mail: dkirkman@ucsd.edu}
                          \thanks{\keckthanks},
  David Tytler\footnotemark[2]
  Nao Suzuki\footnotemark[2],
  Carl Melis, 
  Susan Hollywood, \newauthor
  Kory James,
  Geoffrey So,
  Dan Lubin\footnotemark[2]
  Tridivesh Jena,
  Michael L. Norman, \newauthor
and Pascal Paschos\\
  Center for Astrophysics and Space Sciences, 
  University of California San Diego,
  La Jolla, CA, 92093-0424
}

\usepackage{color}
\usepackage{graphicx}
 
\newcommand{\lya}{\mbox{Ly$\alpha$}}
\newcommand{\lyb}{\mbox{Ly$\beta$}}

\newcommand{\kms}{\mbox{km s$^{-1}$}}
\newcommand{\cmm}{\mbox{cm$^{-2}$}}

\newcommand{\lamr}{\mbox{$\lambda_r$}}
\newcommand{\lamo}{\mbox{$\lambda_o$}}

\newcommand{\ob}{\mbox{$\Omega_b$}}

\newcommand{\om}{\mbox{$\Omega_m$}}
\newcommand{\ol}{\mbox{$\Omega_{\Lambda } $}}

\newcommand{\zabs}{\mbox{$z_{\rm abs}$}}
\newcommand{\zem}{\mbox{$z_{\rm em}$}}

\newcommand{\nhi}{\mbox{N$_{\rm H I}$}}
\newcommand{\lnhi}{\mbox{log \nhi}}

\newcommand{\taueff}{\mbox{$\tau _{\rm eff}$}}

\newcommand{\etal}{{\it et al.}}
\newcommand{\lyaf} {\lya\ forest}

\newif\ifdraftmodep
\draftmodepfalse

\newif\ifapjp
\apjpfalse

\newcommand{\keckthanks}{Visiting Astronomer, W.M. Keck Observatory, which is 
    a joint facility of the University of California, the California
    Instititute of Technology, and NASA}

\begin{document}

\date{\today}
 
\maketitle

\begin{abstract}

We use high quality echelle spectra of 24 QSOs to provide a calibrated
measurement of the total amount of \lyaf\ absorption (DA) over the
redshift range $2.2 < z < 3.2$.  Our measurement of DA excludes
absorption from metal lines or the \lya\ lines of Lyman limit systems
and damped \lya\ systems.  We use artificial spectra with realistic
flux calibration errors to show that we are able to place continuum
levels that are accurate to better than 1\%.  When we combine our
results with our previous results between $1.6 < z < 2.2$, we find
that the redshift evolution of DA is well described over $1.6 < z <
3.2$ as $A (1+z)^\gamma$, where $A = 0.0062$ and $\gamma = 2.75$.  We
detect no significant deviations from a smooth power law evolution
over the redshift range studied.  We find less H~I absorption than
expected at $z=3$, implying that the UV background is about 40\%
higher than expected.  Our data appears to be consistent with an H~I
ionization rate of $\Gamma \sim 1.4 \times 10^{-12} \ \rm{s}^{-1}$.

\end{abstract}


\section{Introduction}

The mean optical depth of the H~I \lyaf\ observed in the spectra of
high redshift QSOs is one of the main pieces of information on the
physical state of the intergalactic medium (IGM).  This is because the
optical depth of the \lyaf\ (\taueff) is sensitive to the combination
of a wide variety of effects that we can collect under two main
headings \citep{rauch97}.  First, \taueff\ is sensitive to all the
familiar parameters of the cosmological model, including \ol , \om ,
\ob, the Hubble parameter and the parameters that describe the
primordial power spectrum of density fluctuations. These parameters
determine the density of hydrogen per unit length, the conversion from
Mpc to wavelength in a spectrum, and the spatial variations of the
density.  The second set of parameters are astrophysical, rather than
primordial, and they determine the ionization and thermal state of the
low density hydrogen.  The IGM is highly photoionized by ultraviolet
photons from early stars and AGN.  The \taueff\ is then sensitive to
the evolution of the intensity and spectrum of the UV background, or
UVB. The energy input per photoionization, and the competition between
photoheating and the cooling from the adiabatic expansion together
give the temperature of the low density IGM, an output rather than an
input parameter.

Over the last decade numerical hydrodynamic simulations of the IGM
have steadily improved in accuracy. We can now make artificial QSO
spectra directly from the full hydrodynamical simulations and measure
the statistical properties of the \lyaf\ absorption in those spectra.
As expected, we find that the absorption in the artificial spectra
depends on the complete set of cosmological and astrophysical
parameters.  In \citet[T04b]{tytler04b} and \citet[J05]{jena05a} we
showed how we can choose sets of input parameters for simulations that
give artificial spectra that are statistically equivalent to the
largest and best samples of real spectra.  We can now use the
simulations to decode the IGM.

When we match simulations to real QSO spectra we obtain joint
constraints on the full set of cosmological and astrophysical
parameters that we input to the simulations. We do not obtain
constraints on individual input parameters, except when we fix all the
many other parameters at values obtained from other observations.

We find that the statistical properties of the \lyaf\ are highly
sensitive to many of the input parameters \citep{tytler04b, bolton05a,
jena05a}.  When we compare to calibrated real spectra, we can expect
to obtain joint constraints on sets of parameters that are competitive
with the best measurements from other types of observations.  This
motivates us to improve the accuracy of the measurement of the IGM.

We also find that the comparison between numerical simulations and
real spectra of the \lyaf\ provides the most accurate measurements of
the intensity of the UVB \citep{rauch97, tytler04b, jena05a,
bolton05a}.

We have found that two statistics in particular provide a good summary
of the \lyaf . One is the effective optical depth \taueff , and the other some
measure of the clumping and temperature of the gas, such as the line
width distribution, or the power spectrum of the QSO flux.  In T04 we
measured \taueff\ to high precision over the redshift range $1.6 < z <
2.2$.  The aim of this paper is to extend the redshift range over
which we can make detailed comparisons between simulations and data by
providing a calibrated and precise measurement of the H~I \lyaf\
opacity over the redshift range $2.2 < z < 3.2$.

\subsection{Previous Work}

There has been extensive previous work dedicated to measuring the
total amount of absorption in the \lyaf , summarized in part by T04b
and \citet{meiksin04a} and the references therein.  Many of these
measurements appear to differ, but \citet{meiksin04a} showed that
much, but not all, of the disagreement between some measurements was
caused by differences in treating errors. T04b discuss other
differences that remain.

Although we have had spectra of the \lyaf\ since 1972, it is only
recently that we have had simulations of the quality to match highly
accurate measurements of the mean amount of absorption.

The mean absorption is hard to measure for three reasons discussed at
length in T04b: the continuum level, metal lines and sample size.

To measure the amount of absorption we must first guess the continuum
prior to the absorption. This is relatively easy at $z \simeq 2$, hard
at $z \simeq 3$ and very hard at $z > 4$ where there is little if any
unabsorbed continuum remaining in the \lyaf .  T04b dealt with the
continuum level by making and fitting artificial spectra that looked
similar to the real spectra. They saw and fitted the emission lines in
the \lyaf\ of each individual spectrum \citep{tytler04a}.  These lines
vary a lot from QSO to QSO \citep{tytler04a, suzuki05a}.  Using the
artificial spectra we were able to show that their continuum level was
accurate to approximately 0.3\%, after correction, and averaging over
the \lyaf\ of 77 QSOs.

The simulations that we compare to the \lyaf\ spectra typically lack
the resolution and physics required to give realistic metal absorption
lines and \lya\ lines from regions with column densities \lnhi $>17.2$
\cmm\ We then need a prescription for dealing with metal lines and
strong \lya\ lines in the \lyaf .  Some measurements ignore these
lines, while others subtract some or most of them.  At $z=1.9$ T04b
showed that the metal lines contributed 15\% and strong \lya\ lines
7\% of the absorption in the \lyaf . However, they each contributed
about the same amount to the total variance of the absorption as did
the \lya\ absorption in the low density IGM.

The last requirement for an accurate measurement of the mean amount of
absorption in the \lyaf\ is a large sample, ideally at least tens of
QSOs.  It had long been noted that there is conspicuous variation in
the amount of \lyaf\ absorption from QSO to QSO \citep{carswell82,
kim01}, and T04b showed that the amount of variation on scales of
$\Delta z = 0.1$ (121 \AA\ in the observed frame) is consistent with
large scale structure for a primordial spectrum of perturbations with
slope $n = 0.95$, and present amplitude $\sigma_8 = 0.9$

Following \citet{oke82} we define ${\rm DA}(z) = 1 - \overline{F}(z)$,
where ${F}(z)$ is the observed flux divided by the continuum level,
and $\overline{F}(z)$ is the mean over many spectra at a given
redshift.  T04b found that $DA(z=1.9) = 0.151 \pm 0.006$ including all
absorption at rest frame wavelengths 1070 -- 1170~\AA\ towards 77
QSOs. The error here is partly from the continuum level, and partly
from the sample size.  T04b estimated the metal line absorption from
wavelengths between the \lya\ and C~IV emission lines, from both their
own spectra and from spectra of \citet{sargent88a}.  They estimated
the strong \lya\ lines from the statistics of such lines in other
spectra.  When they subtracted both the metal lines and strong \lya\
lines the DA drops to $0.118 \pm 0.010$.  T04b estimated that
approximately 5 ideal spectra, all at the same \zem\ with no continuum
errors, and no metal lines or strong \lya\ lines would give DA with an
error of 0.01 at a single redshift, $z=1.9$.  High resolution spectra
might approach this limit.

In \citet{jena05a} we presented a set of 40 fully hydrodynamic
simulations of the IGM at $z=2$. We derived scaling laws that related
the parameters of simulated spectra to the parameters that we input to
the simulations.  When we apply the scaling laws to a simulation, we
can predict the output parameters to higher accuracy than the typical
measurement error in real spectra.  We were able to predict the most
common line width ($b$-value) to 0.3~\kms\ and the \taueff\ to 0.0027,
both approximately a factor of four smaller than the measurement
errors in the real spectra.  In this paper we address the need for
improved measurements.

\subsection{Our Approach}

We apply the methods of T04b to make a calibrated measurement of DA at
$2.2 < z < 3.2$.  The basic idea is that we will ensure that our
continuum fitting is unbiased by simultaneously fitting our real data
and artificial spectra that have been carefully prepared to exhibit
the types of errors shown in real spectra.  The hope is that any
systematic errors in our continuum fitting will manifest themselves in
our continuum fits to both the real data and the artificial data.  We
can then measure them in the artificial data and apply the appropriate
corrections to our real data.

We will generally follow the details of T04b, with some specific
exceptions.  Here we made artificial continuum and emission line
spectra using principal component spectra, rather than real HST
spectra \citep{suzuki05a}. We made the \lyaf\ absorption from randomly
placed Voigt profiles with parameter distribution functions taken from
the literature, instead of using a simplified model of the IGM to
produce the \lyaf\ absorption.  In addition, we added metal and strong
\lya\ lines to the artificial spectra.  However the main difference is
that we now use HIRES spectra with 8~\kms\ resolution in place of the
250~\kms\ resolution spectra that we used in T04b.  The higher
resolution comes with many times more photons per \AA\ and allows us
to place accurate continua at higher redshifts.

\section{Data Sample}

We use a collection of QSO spectra obtained with the HIRES
spectrograph on the Keck telescope \citep{vogt94a}.  These spectra
were collected between 1994 and 2004, for a variety of programs.  The
24 QSOs that we use in this paper were selected from among our HIRES
spectra for the following reasons: (1) they have significant coverage
for \lya\ lines with $2.2 < z < 3.2$, (2) they have SNR 8 -- 70 per
2.1 ~\kms , with a mean of 20 and (3) we were able to perform high
quality flux calibration of the QSO spectra.  The list of spectra we
use in this paper is given in Table 1.

The spectra were all obtained with the original Tektronix 2048 x 2048
24 micron pixel CCD in HIRES up until August 2004.  We used Tom
Barlow's makee package to extract the CCD images and apply the
wavelength calibration.  The individual exposures were flux calibrated
and combined via the procedure described in \citet{suzuki03b}.  We
measured the redshift of each QSO from the estimated peak of the \lya\
emission line, which is given in Table \ref{tabdat} as $z_{\rm em}$
(e.g. $z_{\rm em}$ is not the systemic redshift of the QSO).

\begin{figure}
  \includegraphics[width=84mm]{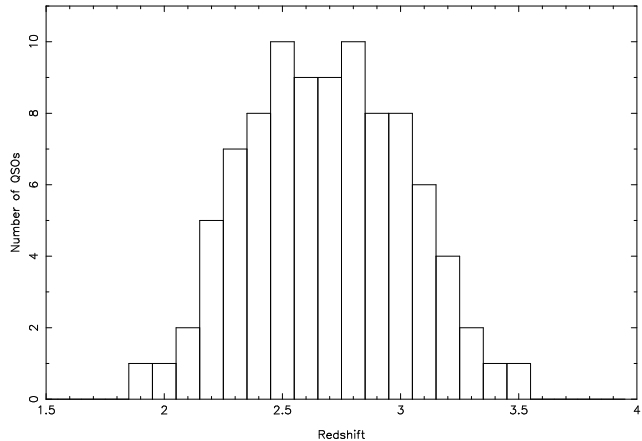}
  \caption{The number of HIRES QSOs which have complete
           \lyaf\ coverage in different redshift bins}
  \label{figzdist}
\end{figure}

A histogram of the number of different QSOs that contribute to each
$\Delta z = 0.1$ redshift bin is shown in Figure \ref{figzdist}.

\begin{table}
\caption{HIRES \lyaf\ spectra used to measure DA}
\label{tabdat}
\begin{tabular}{lllll}
\hline
Identifier  &    RA (B1950)    &      Dec (B1950)      &    $z_{\rm em}$  &  V\\
\hline
q0014+8118 & 00 14 04.45 & +81 18 28.6 & 3.366 & 16.50 \\
q0042-2627 & 00 42 06.42 & -26 27 45.3 & 3.289 & 18.47 \\
q0100+1300 & 01 00 33.38 & +13 00 12.1 & 2.681 & 16.57 \\
q0105+1619 & 01 05 26.97 & +16 19 50.1 & 2.640 & 16.90\\
q0119+1432 & 01 19 16.21 & +14 32 43.2 & 2.870 & 16.70 \\
q0130-4021 & 01 30 50.28 & $-$40 21 51.0 & 3.023 & 17.02\\
q0139+0008 & 01 39 40.85 & +00 08 17.8 & 3.405 & 18.32 \\
q0301$-$0035 & 03 01 07.70 & $-$00 35 03.0 & 3.231 & 17.62 \\
q0322$-$3213 & 03 22 11.18 & $-$32 13 34.6 & 3.302 & 17.95 \\
q0450$-$1310 & 04 50 54.00 & $-$13 10 39.0 & 2.300 & 16.50 \\
q0449$-$1645 & 04 49 59.00 & $-$16 45 09.0 & 2.677 & 17.00 \\
q0741+4741 & 07 41 42.05 & +47 41 53.4 & 3.210 & 17.50\\
q0930+2858 & 09 30 41.40 & +28 58 53.0 & 3.428 & 17.50 \\
q1005+3638 & 10 05 44.13 & +36 38 02.4 & 3.125 & 17.85 \\
q1155+2640 & 11 55 07.62 & +26 40 37.0 & 2.850 & 17.60 \\
q1200+1539 & 12 00 57.62 & +15 39 36.1 & 2.981 & 17.10 \\
q1208+1011 & 12 08 23.81 & +10 11 08.5 & 3.822 & 17.90 \\
q1243+3047 & 12 43 44.90 & +30 47 54.0 & 2.560 & 17.00\\
q1244+3143 & 12 44 48.83 & +31 43 02.9 & 2.961 & 17.90 \\
q1320+3927 & 13 20 41.02 & +39 27 46.8 & 2.985 & 17.06 \\
q1337+2123 & 13 37 47.92 & +21 23 54.1 & 2.700 & 17.90 \\
q2223+2024 & 22 23 13.32 & +20 24 58.5 & 3.560 & 18.38 \\
q2337+1845 & 23 37 13.08 & +18 45 12.2 & 2.620 & 17.00\\
q2344+1229 & 23 44 13.2  & +12 28 50   & 2.784 & 17.5 \\
\hline
\end{tabular}
\end{table}

\section{Artificial Spectra}
We attempted to make artificial spectra that mimic real spectra in
every important way, including emission lines, cosmic ray hits,
echelle blaze, and flux calibration errors.  For each real spectrum,
we made four artificial spectra with the same redshift and noise
properties and various absorption lines.

We made the unabsorbed continuum shape, including emission lines with
the principal component spectra described by \citet{suzuki05a}.  The
artificial spectra have a wide variety of shapes, and include
realistic emission lines between \lya\ and \lyb .  The exact shape and
strength of the emission lines was different for each artificial
spectrum.

We made the H~I absorption from discrete lines, with the line
distribution functions given in \citet{kirkman97a}.  The artificial
spectra used in this paper have lines of all column densities,
$10^{10} < \nhi < 10^{18}$ \cmm, and we placed lines at random
redshifts, from the \zem\ listed in Table 1, down to zero redshift.

If a real spectrum had a DLA with \lnhi $> 10^{19.5}$ ~\cmm , we added
a line with the same \lnhi\ at the same wavelength to all the
accompanying artificial spectra.

With each H~I absorber, we also added the strong doublet absorption
lines of C~IV, Si~IV, and Mg~II to the artificial spectrum.  We held
the H~I/X column density ratio constant for each metal ion, and the
width of each metal absorber was calculated by assuming that the H~I
$b$ values were entirely thermal.  This resulted in an artificial metal
forest that was superimposed on the H~I forest, which made the
artificial spectra look more realistic when inspected closely.

The artificial spectra used in this paper were generated differently
than the spectra we used for the same purpose in T04b.  The spectra in
T04b were generated via a toy model of the IGM that mimicked the large
scale structure, but we did not add metal lines of the \lya\ lines of
Lyman limit systems.

We added noise to the artificial spectrum, and attempted to match the
approximate SNR level of each real spectrum.  We based the noise
levels in our simulated spectra upon the estimated error estimates in
the real echelle spectra.  The resulting artificial spectra have SNR
similar to their real spectra except for three of the five with SNR
$>45$, where the artificial have a SNR too low by approximately
1.5.

\begin{figure}
  \includegraphics[width=84mm]{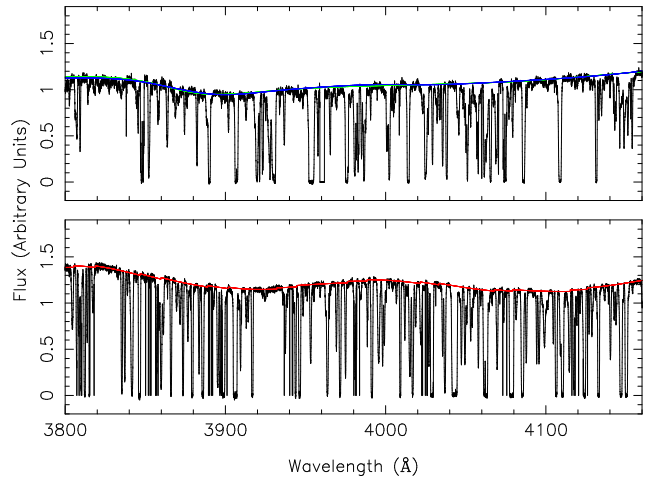}
  \caption{Example of our data, our continuum fits, and the artificial
  spectra we used to verify our continuum fitting ability.  The top
  panel shows the spectrum of Q1243+3047 along with the individual
  continuum fits of our two best fitters.  The bottom panel shows one
  of the four artificial spectra that we made similar to the
  Q1243+3047 spectrum.  Q1243+3047 is representative of the spectra at
  the top end of the SNR distribution for spectra used in this
  paper. The lowest major tick mark is the zero flux level. We show
  rest wavelengths 1070 -- 1170 ~\AA .}
  \label{figsimfit}
\end{figure}

We also added a blaze effect to the artificial noise levels to
simulate the increased SNR obtained at the center of each echelle
order.  Furthermore, we added small fluxing errors, by offsetting
different echelle orders by the blaze function multiplied by a
Gaussian random deviate with $\sigma$ = 2\%, which is larger than the
expected fluxing errors in the data \citep{suzuki03b}.  In Figure
\ref{figsimfit} we show an example of artificial spectra, below the
real spectra they approximate.

Although we see differences between our artificial and real spectra,
they do not concern us.  Since we placed the \lya\ lines at random
redshifts, the artificial spectra lack large scale structure. The
artificial spectra also show more total absorption than real spectra,
generally about 5\% more total absorption over the redshift range $2.4
< z < 3.0$.  As we show in the next section, we are able to fit
continua very well for a wide range of total absorption, so we do not
believe that the minor differences between our real spectra and the
artificial ones have significantly changed our results.

\section{Continuum Fitting}
\label{seccontin}

Four undergraduate authors, SH, KJ, CM, and GS, took a training
program to fit QSO continua.  They fit artificial QSO spectra
that were not matched to any specific QSO.  These spectra had a
variety of emission redshifts, SNR, emission line profiles, and flux
calibration errors. The four all fit the same spectra, and after they
completed a few spectra, we revealed the true continuum level and we
discussed their fit. Within a few weeks two of them were able to fit
continua as well as any of us.

After this training completed, we asked them to fit the real and
matched artificial spectra.  We provided spectra in sets comprising
one real spectrum and two artificial ones matched to that real
one. Each fitter was given the same copy of the real spectrum, but
different versions of the artificial spectra.  We did not reveal which
was the real spectrum in a set.

The results indicate that the two best fitters were excellent.  The
other two fitters were less accurate, and more importantly,
occasionally had made large fitting errors.  Thus for the measurement
of DA, we used the average continuum of the two best fitters, and do
not discuss the results of the other two fitters any further.

The standard deviation of the continuum fit errors per spectral
segments of length $\Delta z = 0.1$ is 1.2\%.  We fit a total of 96
artificial spectra, four per QSO (two per fitter per QSO), and there
were a total of 275 segments of length $\Delta z = 0.1$ in these
artificial spectra. Averaged over all 275 segments, the continua of
the two best fitters were above the true continua by 0.29 \%.  If the
fits were unbiased, and the errors per segment were uncorrelated we
would expect the bias to be $1.2 / \sqrt{275} = 0.07$ \%. As in T04b we
have measured a small bias.

\begin{figure}
  \includegraphics[width=84mm]{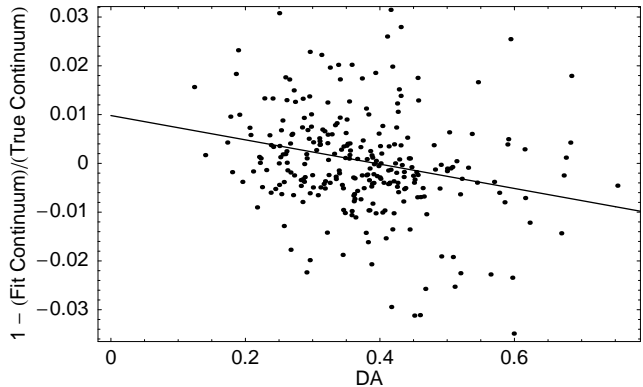}
  \caption{The difference between the fit continuum level and the true
           continuum level for each $\Delta z = 0.1$ segment in the
           artificial spectra.  We plot this difference against
           the mean DA in each spectral segment.  There is a weak
           trend with the amount of absorption, which we fit with the
           solid line.  There are more points on this plot than we
           have in our real data, because we fit four artificial
           spectra for each real spectrum in our analysis.  Note that
           the magnitude of the effect is quite small.}
  \label{figbias}
\end{figure}

In Figure \ref{figbias} we show a weak correlation between the error
in the continuum fit to the artificial spectra and the mean amount of
absorption in a spectral segment of length $\Delta z = 0.1$.  We fit
this correlation as bias $B = 0.0098 - 0.0249 DA$ and we corrected all
our real continua using this fit to remove the trend.  After this
correction when we average over all artificial spectra there is no
remaining bias. All the data on the real HIRES spectra that we give in
the paper have had this correction applied.  We followed a similar
procedure in T04b.  After removing the bias associated with the mean
flux of each spectral segment, we see no correlation between the error
in the continuum fit to the artificial spectra and their SNR.  We have
not measured whether the bias varies from object to object, or as a
function of $z$.

Based upon our ability to continuum fit artificial spectra, we
conclude that errors in continuum placement are not a significant
source of error when measuring DA in high resolution, moderate SNR
spectra at $z < 3.0$.

\section{Measurement of the Mean Flux}

To avoid confusion between multiple Lyman series lines, we restrict
our analysis to the region between the \lya\ and \lyb\ emission lines
of each QSO.  To avoid continuum fitting problems associated with
rapidly changing emission line profiles, and possible contamination
from the proximity effect, we further restrict our measurement of DA to
the rest frame wavelengths 1070 -- 1170 \AA.

\subsection{Removal of LLS absorption}

Because many of our objects were originally observed as part of our
program to measure the primordial deuterium abundance, which can only
be measured in LLS (including DLAs), our data contains more LLS
absorption than would a completely random sample.  This prevents us
from removing the LLS absorption statistically, as we did in T04b.  To
overcome this problem, we have masked out all regions of each spectrum
containing \lya\ absorption associated with identified LLS.  By mask,
we mean that the pixels were marked as being unusable, and were not
used in any further computations.  In practice, our LLS
identifications are probably not complete, but we believe that we
identified and masked all absorbers with $\nhi > 10^{19} \cmm$.  

If the QSO was originally observed as part of our primordial D/H
program, we also masked out any metal lines that might be at the known
redshift of the targeted LLS.  All wavelengths within 2000 \kms\ of an
H~I, C~IV, Si~IV, C~III, SiIII, C~II, or SiII transition at the D
redshift were masked.  The transitions were masked based solely on the
redshift, the transition rest wavelength, and the fact that there is
metal absorption associated with the system in which we had searched for D. 
No attempt was made to fit the metal absorption.

\subsection{Removal of metal line absorption}
\label{metalsec}

We have not attempted to identify and mask individual metal lines in
the \lyaf.  While many of the spectra used in this paper have enough
SNR to identify some of the absorption in the \lyaf, manual removal of
metal absorption is likely to be very incomplete, and we will not know
how much absorption was missed.  Instead, we will use the method
described in T04b, and subtract the metal absorption statistically.
In this section we use the notation from T04b, summarized in Table 1
of that paper.  Briefly, DM refers to the amount of absorption from
metal lines alone, and DM4 refers to the DM from metal lines listed in
Table 3 of SBS88 averaged into bins of size $\Delta z = 0.1$.

We have measured the amount of absorption from metal lines in spectra
published for 52 QSOs in \citet[SBS88]{sargent88a}.  We previously did
this in T04b \S 8.1 for 26 QSOs with 1.7 $<$ \zem\ $< 2.3$. We now add
the remaining QSOs listed in SBS88, to cover $1.7 <$ \zem\ $< 3.54$.
We sum the equivalent widths of all the metal lines listed for the 52
QSOs in Table 3 of SBS88 from rest wavelengths 1225 -- 1500 ~\AA . We
group the lines in bins of length 121.567~\AA\ in the observed frame,
corresponding to $\Delta z = 0.1$ for \lya . In T04b we called these
DM4 values.  We now have 354 DM4 values, with a mean of $0.0191 \pm
0.0017$ and $\sigma (DM4) = 0.0312$, both slightly larger than the
values (mean $0.0167 \pm 0.0022$, $\sigma = 0.0274$) in T04b.  The
mean observed wavelength is now 4764.4~\AA , compared to 4124.8~\AA\
in T04b.

We attempted to remove all absorption lines from systems that appeared
to be associated with the QSOs.  These systems produce large BAL like
lines that are concentrated in certain emission lines, including
Si~IV, N~V and O~VI.  The Si~IV and especially the N~V lines can
contribute a lot to the DM estimate.  However, the lines of these
associated systems will have little effect on the total metal
absorption in the DA region because O~VI is excluded. We found six
QSOs with conspicuous strong associated N~V absorption that was
responsible for many of the largest DM measurements of any segments in
the sample. We also removed all lines from systems with $\beta = v/c <
0.01$, corresponding to velocities within 3000~\kms\ of the QSO.  This
value is large enough to remove most associated systems, without
removing too many intervening systems.  This criterion removed Si~IV
from 6 additional QSOs, and a few C~II and Si~II lines.  We also
removed the occasional Galactic Ca~II at zero redshift.  After
removing these absorption lines, the mean DM decreased 18\% to
$0.0157 \pm 0.0013$ and the standard deviation decreased 21\% to
$\sigma (DM4) = 0.0248$.

Following T04b, we re-fit the DM4 values as a function of 
wavelength in the rest frame of the QSO \lamr\ and observed wavelength
\lamo
\begin{equation}
\label{eqdmfive}              
DM5(\lambda _ r) = 0.01564 - 4.646 \times 10^{-5}  (\lambda _r -1360),
\end{equation} 
and
\begin{equation}
\label{eqdmsix}
DM6(\lambda _ 0) = 0.01686 - 1.798 \times 10^{-6} (\lambda _o -4158).
\end{equation}
(The slope of $DM6(\lambda _ 0)$ was erroneously listed as $7.136
\times 10^{-5}$\% in T04b Eqn. 13, it should be $7.136 \times
10^{-4}\%$).  These fits supersede those given in T04b, both because
the sample is twice as large, and because we did not remove associated
systems in T04b.  Note that here we give the DM as absolute values,
while in T04b we gave the DM values in percentages.

The removal of the associated systems has no significant effect on
the intercepts, but a large effect on the slopes, because the N~V
lines concentrate near the minimum \lamr\ that we used from SBS88.
Removing the associated systems without enlarging the sample causes a
slight decrease in the slope with \lamr\ while increasing the sample has
no additional effect.  For \lamo\ half the decrease in the slope comes
from the removal of the associated systems and the other half from the
enlargement of the sample.

We want DM as a function of both \lamo\ and \lamr .  We see a clear
trend of DM increasing as \lamr\ decreases, but no significant trend
with \lamo , and hence no strong evolution with \zem\ and \zabs .  The
tendency for DM to rise with falling \lamr\ is hard to see in plots,
but appears to come from a smaller fraction of segments with no
absorption lines at lower \lamr .

We choose to correct each bin, from Kast and HIRES, by the DM for its
\lamo\ assuming it is at \lamr = 1120~\AA .  The DM values range from
$DM(1070)= 0.0291$, to $DM(1120)= 0.0268$ in the center of our \lamr\
range, to $DM(1170)= 0.0245$. In T04b we used $DM3 = 0.023 \pm 0.005$
for all segments.  We now use $DM(1120)= 0.0268$, for all segments,
and we simultaneously correct for the slight trend with \lamo\ given
by Eqn. \ref{eqdmsix}.  At the mean \lamo\ of the Kast sample we
measure $DM(3525) = 0.01799$ while for the HIRES sample $DM(4498) =
0.01624$, both at the mean \lamr\ of the SBS88 sample: 1357.2~\AA .
We apply the \lamo\ correction by multiplying by \taueff (\lamo
)/\taueff (4764.4~\AA ), where 4764.4~\AA\ is the mean \lamo\ of the
SBSS88 sample. For example, we calculate \taueff\ at 4498~\AA , to be
$1.03 \times 0.02715$ corresponding to $DM = 0.0276$.  As we explained
in \citet{jena05a}, we subtract the \taueff\ values corresponding to
the DM from the \taueff\ corresponding to the DA. We do not directly
subtract $DA-DM$.

Since the DM values that we use are now larger than in T04b, the DA
values are smaller, and the changes are most at the lowest \zabs .

Although we have improved upon T04b, we do not have a definitive
measurement of DM. Three issues remain.  First, we extrapolate in
\lamr\ well beyond the limit of the data.  Second, we use all the
metal lines listed in SBS88, without regard for the equivalent
width. Their SNR was approximately constant with wavelength in the
relevant wavelength range, other than an increase in the N~V emission
line. Since their spectra do not have the resolution or SNR to see
weak lines, the DM values we obtain from their spectra will be too
small, and our DA too large.  Third, our masking the lines of LLS may
leave too few metal lines compared to the DM from SBS88. This error
would make our DA too small.  We masked all the metals from one LLS in
most QSOs. These QSOs were selected because they contained these LLS,
and hence we expect that they would have excess metals if we did not
apply this mask. However, some of a random sample of QSOs would have
LLS by chance, and hence we may have over corrected. This error will
tend to cancel that from the numerous weak metal lines missed from the
SBS88 line lists.

As a sanity check, we measured the amount of (unmasked) metal
absorption in the six of our spectra that have significant coverage
redward of the \lya\ emission line.  In those six spectra we measured
the mean metal absorption over the observed wavelengths that
correspond to the \lyaf\ between $2.2 < z < 3.2$ to be 2.1\%.  This
value likely has large errors associated with it because it has been
measured from only six QSOs, but it does increase our confidence the
our general metal line removal based on the results of SBS88 is not
wildly inaccurate.

\subsection{Measurement of mean DA in $\Delta z = 0.1$ spectral segments}

After masking targeted Deuterium systems and LLS with obvious \lya\
lines, we measured the mean flux in spectral segments of $\Delta z =
0.1$.  We started at a rest wavelength of 1070 \AA\ (or the lowest
wavelength with data in cases of incomplete spectral coverage) in each
QSO, and computed the mean flux of all segments that are fully
contained between the rest wavelengths of 1070 and 1170 \AA.  There
are typically three $\Delta z = 0.1$ segments in an individual QSO
spectrum.  We discard incomplete segments and any segment with more
than 10\% of it's flux masked for any reason, e.g. a DLA in a spectrum
will cause at least one, and frequently two segments in a given QSO
spectrum to be discarded. Finally, we subtracted the anticipated metal
absorption from each segment, as described in \S \ref{metalsec}.

\begin{figure}
  \includegraphics[width=84mm]{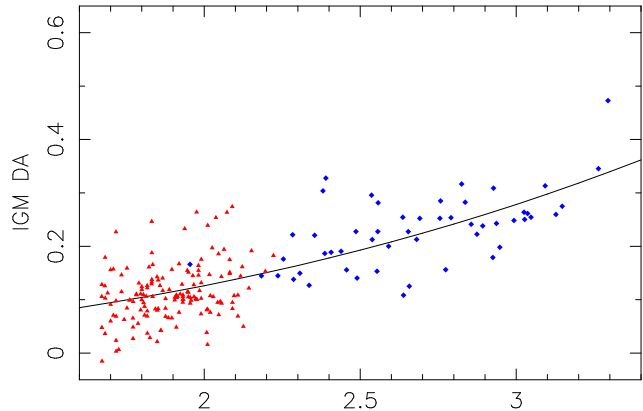}
  \caption{DA as a function of redshift.  Each point shows the DA
  measured along a particular line of sight over a path of length
  $\Delta z = 0.1$.  The light triangles are the data from 77 QSOs used
  by T04b (observed total absorption, minus mean absorption by \lya\ of
  LLS and metals), while the dark diamonds show the new data from 24
  QSOs from this paper (excluding individual \lya\ of LLS, and minus
  the mean metal absorption).  This plot does not show part of the
  data at the higher redshift end of each spectrum because we do not
  show bins that are partly sampled by a spectrum.}
  \label{figkastda}
\end{figure}

Our results are shown in Figure \ref{figkastda}, where we also show
the results for the Kast spectra from T04b.  Figure \ref{figkastda}
shows only the absorption of the low column density \lyaf\ -- we have
attempted to remove the \lya\ of all LLS and 
all metal line absorption for all the Kast
and HIRES spectra.  For both the HIRES and Kast spectra we subtracted the
same mean amount of metal absorption from each Kast point.
For the \lya\ of the
LLS we subtracted the mean for all the Kast points, but we masked 
individual \lya\ lines for the HIRES points.

\subsection{DA as a function of redshift}

\begin{figure}
  \includegraphics[width=84mm]{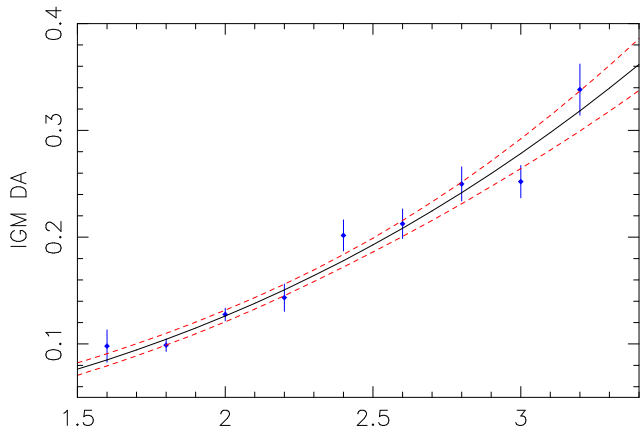}
  \caption{DA as a function of redshift.  Here we have binned the data
  points shown in Figure \ref{figkastda} into bins of $\Delta z =
  0.2$.  The solid line shows the minimum $\chi^2$ fit of the function
  $A (1+z)^\gamma$.  We find $A = 0.0062$ and $\gamma = 2.75$ give
  $\chi^2 = 8.69$ for seven degrees of freedom.  The dashed lines show
  the $\pm 1 \sigma$ confidence interval on the fit.}
\label{figdavsz}
\end{figure}

To tabulate the mean DA as a function of redshift, we binned the data
shown in Figure \ref{figkastda} into redshift bins of width $\Delta z
= 0.2$.  We estimated the mean DA of each bin to be simply the mean
value of the points in the bin.  We estimated the error to be the
standard deviation of {\it all} the points (see \S \ref{secdisp}, we
compute the standard deviation from the best power law fit to the
data) divided by the square root of the number of points in each bin.
Using the standard deviation of all points instead of the standard
deviation of just the points in each bin gives nearly identical
results for bins with large numbers of points, but seems to be much
better behaved for bins with small numbers of points.  Our results are
in Table \ref{tabda} and shown in Figure \ref{figdavsz}.

\begin{table}
\caption{Measured IGM DA values in $\Delta z = 0.2$ redshift bins}
\label{tabda}
\begin{tabular}{lll|ll}
\hline
$z$      &           &                      &   Fit   &   Fit\\
center   &    DA     &   $\sigma_{\rm DA}$  &   DA    &   $\sigma_{\rm DA}$ \\
\hline
1.6 & 0.098 & 0.015 & 0.0851 &  0.0057 \\
1.8 & 0.099 & 0.006 & 0.1044 &  0.0055 \\
2.0 & 0.128 & 0.006 & 0.1262 &  0.0055 \\
2.2 & 0.143 & 0.013 & 0.1507 &  0.0054 \\
2.4 & 0.202 & 0.014 & 0.1780 &  0.0059 \\
2.6 & 0.213 & 0.014 & 0.2083 &  0.0074 \\
2.8 & 0.250 & 0.016 & 0.2417 &  0.0101 \\
3.0 & 0.252 & 0.015 & 0.2783 &  0.0139 \\
3.2 & 0.338 & 0.024 & 0.3183 &  0.0187 \\
\hline
\end{tabular}
\end{table}

DA as a function of $z$ is well fit by a power law of the form $A
(1+z)^\gamma$.  Minimizing $\chi^2$ to the points in Table \ref{tabda}
gives $A = 0.0062$ and $\gamma = 2.75$.  The $\chi^2$ of the best fit
is 8.69 and the reduced $\chi^2$ for 7 degrees of freedom (9 data
points - 2 parameters) is 1.24.

The dashed lines in Figure \ref{figkastda} enclose the $\pm 1 \sigma$
confidence interval fits.  They are produced by taking the envelope
that contains all power law fits with $\chi^2 < \chi^2_{\rm min} +
2.3$.  Thus the plotted error bounds are not power laws, but each
point on the error curves corresponds to a point on a power law that
fits our data.  The last column in Table \ref{tabda} gives the
difference between the bounds divided by two.

We do not give errors for $A$ or for $\gamma$.  The $A$ -- $\gamma$
$\chi^2$ manifold is complex, and it is not well approximated by only
two numbers ($\sigma$ for $A$ and $\gamma$).  Attempts to do so in the
past \citep{steidel87a, press93a, kim01}, have lead to all sorts of
confusion in the literature -- consider the efforts of
\citet{seljak03} and \citet{meiksin04a} to determine the allowed
values of \taueff\ at various redshifts.  We recommend use of the
results given in Table \ref{tabda}.

Note that the small dip in DA that we observed at $z > 2.2$ in Fig. 22 of
T04b is
not present in the new HIRES data.  The combined HIRES + KAST data
seems to be well fit by a single power law, and no significant
deviations from a power law are present in our data.

\subsection{Dispersion of DA in $\Delta z = 0.1$ spectral segments}
\label{secdisp}

The new points from this work show less dispersion than the points in
T04b.  This is consistent with the fact that we have removed LLS
absorption before calculating the mean flux in each point, while T04b
calculated the mean flux in each point with the LLS present, and then
subtracted the LLS absorption statistically.  While both results will
give the same mean value for the H~I \lyaf\ opacity, the T04b method
will include the substantial dispersion of the LLS absorption.

The standard deviation of the HIRES DA points (each covering $\Delta z
= 0.1$) about the mean given by the power law fit is $\sigma_{\rm DA}
= 5.2^{+0.6}_{-0.4}$ \% over redshifts $2.2 < z < 3.2$.  This
$\sigma_{\rm DA}$ value includes the intrinsic variance of the \lyaf,
the variance of metal line absorption (we removed the mean, not the
individual metal absorption), and the variance of our continuum
fitting errors.  It should not include a significant contribution from
LLS absorption, since we masked the \lya\ absorption associated with
LLS before measuring any DA values.

We can subtract variances (squares of standard deviations) to estimate
the standard deviation of the absorption by \lya\ from the low density
IGM alone. As in T04 \S 9, we work with mean values in segments of
length $\Delta z = 0.1$. We now repeat the calculations given in T04,
considering only the HIRES spectra. We take the standard deviation of
the metal absorption as 3.7 \%, where we have scaled the value 3.1 \%
in Table 4 of T04b by the mean metal absorption which is now DM = 2.76
\%, up from 2.3 \%.  We also use the standard deviation of the
continuum fitting errors 1.2 \% (\S \ref{seccontin}). We find that the
intrinsic variation of the mean amount of absorption in the \lyaf\
over $\Delta z = 0.1$ segments is $\sigma (\Delta z = 0.1) =
3.4^{+1.0}_{-1.2} $ \%.  This value is a mean from the whole $z$
range, centered near $z=2.7$.  At $z=2.7$ the mean total DA per
segment, including metals but not \lya\ lines of LLS, is 0.246, and
after subtracting the metals it drops to 0.2246.

The $\sigma (\Delta z = 0.1)$ value from HIRES spectra agrees with the
value that we calculated in T04b (section 9.4) for Kast spectra:
$\sigma (\Delta z = 0.1) = 3.9 ^{+0.5}_{-0.7}$ \% at $z=1.9$.  The
error is larger now, because we have fewer DA measurements giving a
larger error on the measurement of the standard deviation of the DA
values.  We expect that $\sigma (\Delta z = 0.1)$ will increase with
increasing $z$ because the power spectrum of the flux in the \lyaf\
increases with increasing redshift \citep{croft02b, mcdonald04a}.  We
do not see this, but this may be because our measurement errors are
large and not well determined.  In T04b we did not subtract the
continuum fitting error because it appeared to be small, and was not
well determined. We found contradictory evidence that the continuum
fit error might be large, in which case the $\sigma (\Delta z = 0.1)$
value should be less than the $3.9 ^{+0.5}_{-0.7}$ \%.

We conclude that 24 QSOs is adequate to get the mean DA, but we would
prefer far more QSOs to obtain an accurate measurement of the variance
in the DA.

\section{Discussion}

Our DA results are in general agreement with the values of the
literature summary given in \citet[Table B1]{meiksin04a}.  However, we
find approximately $0.03$ less absorption at all redshifts.  This is
consistent with the fact that we have attempted to measure only the
absorption due to the \lyaf, while \citet{meiksin04a} gave values for
all absorption in the \lyaf\ region of a spectrum.

At $z = 2.2$, our results are also consistent with \citet{schaye03a},
who also attempted to measure only the absorption associated with the
\lyaf.  They only removed metals absorption they could identify
directly in the absorption spectra, so their removal completeness is
unknown, and is probably a function of redshift.  However, we find a
shallower redshift evolution and less absorption at $z = 3.0$.

The best fit value of DA = 0.278 at $z=3.0$ is lower than we expected
to find when we started this work.  In T04b and J05 we developed a
concordance model of the \lyaf\ at $z =1.9$ -- this model is referred
to as model ``A'' in T04b and J05.  Model A uses a uniform UVB with
the shape and redshift evolution due to \citet{haardt01a}, and
displayed graphically in \citet{paschos05a}.  Model A, while giving
the measured value of DA at $z=1.9$, predicts DA = 0.34 at $z=3$.  The
clear implication is that there are more ionizing photons at $z=3$
than predicted by \citet{haardt01a}.  J05 showed that the naive
expectation that \taueff\ scales like the H~I ionization rate is valid
at $z=2$. Assuming that this result holds at $z=3$, and translating DA
into \taueff, we find an H~I ionization rate of $\Gamma_{-12} = 1.3
\pm 0.1$, or $1.3$ times the value predicted by
\citet{haardt01a}. The error comes from propagating the observed error
on DA through the J05 scaling relation.  The real error is likely to
be larger, because there is some disagreement as to the scaling
relationship between $\Gamma$ and \taueff.  While J05 found $\Gamma
\propto \taueff^{-1}$, \citet{bolton05a} found $\Gamma \propto
\taueff^{-1.44}$ at $z=2$.  \citet{bolton05a} also found some change
in redshift, with $\Gamma \propto \taueff^{-1.61}$ at $z=3$.  If we
use the \citet{bolton05a} $z=3$ relationship instead of the J05
scaling relation, we instead find $\Gamma_{-12} = 1.5$.

Irrespective of the precise value of $\Gamma$, this data seems to
indicate that it is larger than expected.  It will be interesting to
try to find the source of these extra photons.

\section*{Acknowledgments}

All of the spectra used in this paper were obtained with the HIRES
spectrograph, we would like to thank those who designed, built and
maintain this amazing resource.  We are grateful to the staff and
observing assistants at the W.M. Keck Observatory, who have
supported our observing for many years.  The W.M. Keck Observatory is
a joint facility of the University of California, the California
Institute of Technology, and NASA. We are grateful to the referee, James
Bolton, for several insightful comments. We thank Tom Barlow, who's makee
package was used to extract all of the HIRES spectra used in this
project.  Former UCSD students Xiao-Ming Fan, Scott Burles, Ed Lever,
and John O'Meara helped to obtain some of the spectra used in this
paper.  SH was funded by the NSF REU program grant to the physics
Dept. at UCSD.  This work was funded in part by NSF grants AST-9900842
AST-9803137 and AST-0098731, and NASA grants NAG5-13113 and STScI grant
HST-AR-10288.01-A.

\bibliographystyle{mn2e}
\bibliography{archive}

\begin{thebibliography}{}

\bibitem[\protect\citeauthoryear{Bolton, Haehnelt, Viel \& Springel}{Bolton
  et~al.}{2005}]{bolton05a}
Bolton J.~S.,  Haehnelt M.~G.,  Viel M.,    Springel V.,  2005, \mnras, 257,
  1178

\bibitem[\protect\citeauthoryear{{Carswell}, {Whelan}, {Smith}, {Boksenberg} \&
  {Tytler}}{{Carswell} et~al.}{1982}]{carswell82}
{Carswell} R.~F.,  {Whelan} J.~A.~J.,  {Smith} M.~G.,  {Boksenberg} A.,
  {Tytler} D.,  1982, \mnras, 198, 91

\bibitem[\protect\citeauthoryear{{Croft}, {Weinberg}, {Bolte}, {Burles},
  {Hernquist}, {Katz}, {Kirkman} \& {Tytler}}{{Croft} et~al.}{2002}]{croft02b}
{Croft} R.~A.~C.,  {Weinberg} D.~H.,  {Bolte} M.,  {Burles} S.,  {Hernquist}
  L.,  {Katz} N.,  {Kirkman} D.,    {Tytler} D.,  2002, \apj, 581, 20

\bibitem[\protect\citeauthoryear{Haardt \& Madau}{Haardt \&
  Madau}{2001}]{haardt01a}
Haardt F.,  Madau P.,  2001, in Clusters of galaxies and the high redshift
  universe observed in X-rays, Recent results of {XMM-Newton and Chandra},
  {XXXVIth Rencontres de Moriond}, {XXIst Moriond Astrophysics Meeting, March
  10-17, 2001, Savoie France}. Edited by {D.M. Neumann \& J.T.T. Van} Modelling
  the {UV/X-ray} cosmic background with {CUBA}

\bibitem[\protect\citeauthoryear{{Jena}, {Norman}, {Tytler}, {Kirkman},
  {Suzuki}, {Chapman}, {Melis}, {So}, {O'Shea}, {Lin}, {Lubin}, {Paschos},
  {Reimers}, {Janknecht} \& {Fechner}}{{Jena} et~al.}{2005}]{jena05a}
{Jena} T.,  {Norman} M.~L.,  {Tytler} D.,  {Kirkman} D.,  {Suzuki} N.,
  {Chapman} A.,  {Melis} C.,  {So} G.,  {O'Shea} B.~W.,  {Lin} W.,  {Lubin} D.,
   {Paschos} P.,  {Reimers} D.,  {Janknecht} E.,    {Fechner} C.,  2005, \mnras
  submitted, astro-ph/0412557

\bibitem[\protect\citeauthoryear{{Kim}, {Cristiani} \& {D'Odorico}}{{Kim}
  et~al.}{2001}]{kim01}
{Kim} T.-S.,  {Cristiani} S.,    {D'Odorico} S.,  2001, \aap, 373, 757

\bibitem[\protect\citeauthoryear{{Kirkman} \& {Tytler}}{{Kirkman} \&
  {Tytler}}{1997}]{kirkman97a}
{Kirkman} D.,  {Tytler} D.,  1997, \apj, 484, 672

\bibitem[\protect\citeauthoryear{{McDonald}, {Seljak}, {Burles}, {Schlegel},
  {Weinberg}, {Shih}, {Schaye}, {Schneider}, {Brinkmann}, {Brunner} \&
  {Fukugita}}{{McDonald} et~al.}{2004}]{mcdonald04a}
{McDonald} P.,  {Seljak} U.,  {Burles} S.,  {Schlegel} D.~J.,  {Weinberg}
  D.~H.,  {Shih} D.,  {Schaye} J.,  {Schneider} D.~P.,  {Brinkmann} J.,
  {Brunner} R.~J.,    {Fukugita} M.,  2004, \apj submitted, astro-ph/0405013

\bibitem[\protect\citeauthoryear{{Meiksin} \& {White}}{{Meiksin} \&
  {White}}{2004}]{meiksin04a}
{Meiksin} A.,  {White} M.,  2004, \mnras , {eprint arXiv:astro-ph/0205387},
  350, 1107

\bibitem[\protect\citeauthoryear{{Oke} \& {Korycansky}}{{Oke} \&
  {Korycansky}}{1982}]{oke82}
{Oke} J.~B.,  {Korycansky} D.~G.,  1982, \apj, 255, 11

\bibitem[\protect\citeauthoryear{Paschos \& Norman}{Paschos \&
  Norman}{2005}]{paschos05a}
Paschos P.,  Norman M.~L.,  2005, \apj \ (submitted, astro-ph/0412244)

\bibitem[\protect\citeauthoryear{{Press}, {Rybicki} \& {Schneider}}{{Press}
  et~al.}{1993}]{press93a}
{Press} W.~H.,  {Rybicki} G.~B.,    {Schneider} D.~P.,  1993, \apj, 414, 64

\bibitem[\protect\citeauthoryear{{Rauch}, {Miralda-Escude}, {Sargent},
  {Barlow}, {Weinberg}, {Hernquist}, {Katz}, {Cen} \& {Ostriker}}{{Rauch}
  et~al.}{1997}]{rauch97}
{Rauch} M.,  {Miralda-Escude} J.,  {Sargent} W.~L.~W.,  {Barlow} T.~A.,
  {Weinberg} D.~H.,  {Hernquist} L.,  {Katz} N.,  {Cen} R.,    {Ostriker}
  J.~P.,  1997, \apj, 489, 7

\bibitem[\protect\citeauthoryear{{Sargent}, {Boksenberg} \&
  {Steidel}}{{Sargent} et~al.}{1988}]{sargent88a}
{Sargent} W.~L.~W.,  {Boksenberg} A.,    {Steidel} C.~C.,  1988, \apjs, 68, 539

\bibitem[\protect\citeauthoryear{Schaye, Aguirre, Kim, Theuns, Rauch \&
  Sargent}{Schaye et~al.}{2003}]{schaye03a}
Schaye J.,  Aguirre A.,  Kim T.-S.,  Theuns T.,  Rauch M.,    Sargent W. L.~W.,
   2003, \apj, 596, 768

\bibitem[\protect\citeauthoryear{{Seljak}, {McDonald} \& {Makarov}}{{Seljak}
  et~al.}{2003}]{seljak03}
{Seljak} U.,  {McDonald} P.,    {Makarov} A.,  2003, \mnras, 342, L79

\bibitem[\protect\citeauthoryear{{Steidel} \& {Sargent}}{{Steidel} \&
  {Sargent}}{1987}]{steidel87a}
{Steidel} C.~C.,  {Sargent} W.~L.~W.,  1987, \apj, 313, 171

\bibitem[\protect\citeauthoryear{{Suzuki}, {Tytler}, {Kirkman}, {O'Meara} \&
  {Lubin}}{{Suzuki} et~al.}{2003}]{suzuki03b}
{Suzuki} N.,  {Tytler} D.,  {Kirkman} D.,  {O'Meara} J.~M.,    {Lubin} D.,
  2003, \pasp, 115, 1050

\bibitem[\protect\citeauthoryear{{Suzuki}, {Tytler}, {Kirkman}, {O'Meara} \&
  {Lubin}}{{Suzuki} et~al.}{2005}]{suzuki05a}
{Suzuki} N.,  {Tytler} D.,  {Kirkman} D.,  {O'Meara} J.~M.,    {Lubin} D.,
  2005, \apj, 618, 592

\bibitem[\protect\citeauthoryear{Tytler, Kirkman, O'Meara, Suzuki, Orin, Lubin,
  Paschos, Jena, Lin \& Norman}{Tytler et~al.}{2004}]{tytler04b}
Tytler D.,  Kirkman D.,  O'Meara J.,  Suzuki N.,  Orin A.,  Lubin D.,  Paschos
  P.,  Jena T.,  Lin W.-C.,    Norman M.,  2004, \apj , astro-ph/0403688, 617,
  1

\bibitem[\protect\citeauthoryear{{Tytler}, {O'Meara}, {Suzuki}, {Kirkman},
  {Lubin} \& {Orin}}{{Tytler} et~al.}{2004}]{tytler04a}
{Tytler} D.,  {O'Meara} J.~M.,  {Suzuki} N.,  {Kirkman} D.,  {Lubin} D.,
  {Orin} A.,  2004, \aj, 128, 1058

\bibitem[\protect\citeauthoryear{{Vogt}, {Allen}, {Bigelow}, {Bresee}, {Brown},
  {Cantrall}, {Conrad} \& {Couture}}{{Vogt} et~al.}{1994}]{vogt94a}
{Vogt} S.~S.,  {Allen} S.~L.,  {Bigelow} B.~C.,  {Bresee} L.,  {Brown} B.,
  {Cantrall} T.,  {Conrad} A.,    {Couture} M.~{\it et al}.,  1994, in Proc.
  SPIE Instrumentation in Astronomy VIII, David L. Crawford; Eric R. Craine;
  Eds., Volume 2198, p. 362 Vol.~2198, {HIRES: the high-resolution echelle
  spectrometer on the Keck 10-m Telescope}.
p.~362

\end{thebibliography}

\end{document}